\documentclass{article}
\usepackage[utf8]{inputenc}
\usepackage{authblk}
\usepackage{graphicx}
\usepackage{geometry}
\usepackage{multirow}
\usepackage{enumerate}
\title{Towards Automated Benchmark Support for Multi-Blockchain Interoperability-Facilitating Platforms}
\date{}
\author[1]{Mostafa Kazemi}
\author[2]{Abbas Yazdinejad}
\affil[1]{Department of Electrical Engineering, Faculty of Engineering, Shahed University, Tehran, Iran (kazemi.mu@gmail.com)}
\affil[2]{School of Computer Science, University of Guelph, Ontario, Canada (ayazdine@uoguelph.ca)}

\begin{document}

\maketitle
\begin{abstract}
Since the introduction of the first Bitcoin blockchain in 2008, different decentralized blockchain systems such as Ethereum, Hyperledger Fabric, and Corda, have emerged with public and private accessibility. It has been widely acknowledged that no single blockchain network will fit all use cases. As a result, we have observed the increasing popularity of multi-blockchain ecosystem in which customers will move toward different blockchains based on their particular requirements. Hence, the efficiency and security requirements of interactions among these heterogeneous blockchains become critical. In realization of this multi-blockchain paradigm, initiatives in building Interoperability-Facilitating Platforms (IFPs) that aim at bridging different blockchains (a.k.a. blockchain interoperability) have come to the fore. Despite current efforts, it is extremely difficult for blockchain customers (organizations, governments, companies) to understand the trade-offs between different IFPs and their suitability for different application domains before adoption. A key reason is due to a lack of fundamental and systematic approaches to assess the variables among different IFPs. To fill this gap, developing new IFP requirements specification and open-source benchmark tools to advance research in distributed, multi-blockchain interoperability, with emphasis on IFP performance and security challenges are required. In this document, we outline a research proposal study to the community to realize this gap. 
\end{abstract}

\section{Introduction}
As a cryptographic-based distributed ledger, blockchain technology \cite{1,9226671} enables trusted transactions among untrusted participants in the network. Most notably, there is an emerging trend beyond cryptocurrency payments, transforming the blockchain into a new paradigm of decentralized systems development \cite{2}. Recently, blockchain technology has been the subject of increased scientific research and development, and has raised significant interest among researchers, developers, and industry practitioners \cite{a3,9163144,9162735,9129462,Srivastava2020,zhang2018pentagon,8946262,TAYLOR2020147,8818320,8946164,8861782,102227,8917991,ekramifard2020systematic,CHEN2020102370}. Since the introduction of the first Bitcoin blockchain in 2008 \cite{3}, various blockchain systems have been developed (such as Ethereum, Hyperledger Fabric, and Corda) and has been widely acknowledged that there will be no single blockchain suitable for all use cases. A more likely scenario is that different customers will move toward other blockchains based on their particular requirements \cite{4}. For example, financial applications requiring consensus between a fixed set of banks are likely to adopt a blockchain such as Corda or Hyperledger Fabric. Hyperledger Fabric is based on the concept of private blockchain which is designed for enterprise customers who usually require high throughput and data privacy. Other applications that may require fully open participation are likely to adopt an open blockchain such as Ethereum, which introduces the concept of smart contracts \cite{5, 6} on top of a blockchain. At the same time, the sidechain technology \cite{7, 8, 9} has most recently emerged as a separate chain attached to the main chain, in parallel with transactions, to alleviate the overhead on the main chain \cite{a2}.  

In the out-years we envision a distributed multi-blockchain ecosystem, in which different blockchains are required to collaborate with each other in various scenarios \cite{a4,8488201}. For example, if Walmart builds its supply chain using Hyperledger Fabric and its finance chain on Corda, the two chains could interact in such a way that a financial transaction could require a confirmation on the Corda in order to trigger the corresponding shipping transaction on Hyperledger Fabric. To achieve the much-discussed potential of distributed multi-blockchain and eliminate (or at least significantly open) the existing landscape of proprietary data silos, it is thus essential to enable  blockchain interoperability that allows these disparate blockchains to communicate with each other. Consequently, the efficient and secure interactions among these heterogeneous blockchain systems become critical.
 
To realize this multi-blockchain paradigm there has been a recent  growth of initiatives \cite{10, 11, 12} to build Interoperability-Facilitating Platforms (IFPs) that aim to bridge different blockchains (a.k.a. blockchain interoperability). The goal of IFPs is to build a decentralized network that allows independent disparate blockchains with varying governance principles to interact and transact with one another. From a blockchain interoperability perspective, however, blockchain customers still lack a clear and practical understanding of such platforms, and many fundamental questions remain unresolved. For example, unresolved issues include: How efficient are these platforms in performing cross-chain data transactions? How do these platforms establish security defaults to prevent attacks? How could IFP customers be assured of the integrity and confidentiality of their data? Without the answer to these research-oriented questions, it will be extremely difficult for blockchain customers (organizations, governments, companies) to understand the trade-offs between different IFPs and their suitability for different application domains before adoption. Therefore, it is critical to build fundamental and systematic approaches to benchmark and assess variables amidst different IFPs. 
 
As new systems and paradigms emerge, new benchmarking tools are required in order to support researchers, organizations, and developers to conduct assessment studies and trade-off analysis \cite{13}. With the exception of Blockbench \cite{14} for private blockchains and Performance Traffic Engine \cite{15} for Hyperledger Fabric blockchain, few blockchain-related benchmarking proposals currently exist. These limited proposals focus mainly on single-blockchains and do not take into consideration the interoperability of multi-blockchain ecosystem. \textit{This project proposal propose to fill this gap by developing new IFP requirements specification and a benchmark tool to advance research in multi-blockchain interoperability, with emphasis on the challenges related to IFP performance and security. The overarching goal is to create a standard multi-blockchain benchmark discipline to assist with the evaluation and improvement of different IFPs.}

To achieve the stated objectives, three major activities are proposed to be performed using DevOps-like Agile methodology \cite{16, 17, 18, 19}, which emphasizes an incremental and evolutionary development approach. This nimble methodology will help to rapidly develop a prototype of IFP benchmarking framework for evaluation and increase the sustainability of the proposed research beyond the life of funding by building and engaging IFP and blockchain research, development, and user communities. 

Specifically, we propose to perform three main tasks, guided by the following inquiries, posited under each task:

\textbf{Task 1: Defining IFP requirements specification and core benchmarks.} The goal of this task is to provide a clear understanding of what each component does in an IFP and how it interfaces with blockchains and the rest of its environment by considering their fundamental properties. To do so, we suggest to critically examine existing IFPs and blockchains to understand which IFP components performance are- critical as well as security sensitive. We will use these insights to develop new requirements specification and core benchmarks for representative IFPs (e.g., ICON \cite{11}, AION \cite{10}, and Wanchain \cite{12}). In our context, a benchmark is defined as a software tool that generates customized cross-chain transactions as client workloads, execute the workloads on the target IFP, monitor the behaviors of the transaction execution, and report the results to the users. The following research questions should be answered in this phase:

\textbf{\textit{RQ1a:}} \textit{How do the existing IFPs work in practice?} Currently, there are various IFPs such as ICON, AION, Wanchain, and BTC Relay \cite{20}. Although they all aim at building connectivity between different blockchains by enabling cross-chain transactions, they utilize different techniques and designs. Hence, it is important to start by thoroughly fleshing out their inner workings.

\textbf{\textit{RQ1b:}} \textit{What are the core requirements that an IFP needs to possess to match the expectations of users who seek interoperability between different blockchains?} As with many blockchain platforms, which offer specifications (e.g., the Enterprise Ethereum Client Specification V2 \cite{21}), we will fundamentally define a set of requirements specification with respect to target quality factors, including performance, security, and usability of an IFP.

\textbf{\textit{RQ1c:}} \textit{What are the characteristics and properties that can effectively define benchmarks for an IFP?} The answers to this question will help us to systematically define and structuralize IFP benchmarks, which will be the building blocks of the proposed benchmark tool. The benchmark tool includes benchmarks that support different IFPs, where each benchmark can generate various workloads based on user inputs to evaluate IFPs. We will develop the core benchmarks in three tiers: IFP functionality related, IFP performance related, and IFP security related. It is worth mentioning that our extensible architecture (discussed in Task 2) will foster the development of user-defined benchmarks on top of the core benchmarks. 

\textbf{Task 2: Building the benchmarking framework with extensible architecture.} The goal of this task is to create an extensible programmable framework with documentation that will support easy definition of additional benchmarks for new IFPs, and, most importantly, will also make it easier to integrate different IFPs via simple APIs. In parallel, and inspired by our Agile methodology, we will build a prototype of the benchmark tool for public use, in both academia and industry, and seek feedback. We will also perform an assessment on the prototype by benchmarking commonly used IFPs, such as ICON, AION, and Wanchain using the proposed core benchmarks in Task 1. This early assessment will help us gain insights into the functionality of our benchmark tool, as it evolves through the project period. The results of this assessment will be published along with the release of the prototype. The following research questions should be addressed in this phase:

\textit{\textbf{RQ2a:} What components are required to build the framework?} Considering our three main benchmark tiers (functionality, performance and security), we will design the major components of the framework, including a client workload generator, client workload executor, transactions behavior monitor, IFP interface layer and a visualization GUI to analyze, understand, and report the results of the benchmarking process. 

\textit{\textbf{RQ2b:} How do we model the framework to make its underlying foundation easily connected to different IFPs?} To address this question, we will use a microservice architecture \cite{22} as our design approach to make the framework extensible. The resulting architecture will horizontally scale and enhance interoperability (i.e., component implementations in this architecture are expected to be easily replaceable). The most important benefit of this approach is that it will ensure the benchmarking framework will include not only a repeatable process, but also a detailed template for formulating a microservice architecture implementation (see Task 3) by any client.
                                               
\textbf{Task 3: Constructing the full-fledged automated GUI-based IFP benchmark tool.} The goal of this task is to use the developed prototype as a basis to provide the automated tool support for the multi-blockchain interoperability benchmarking. Using the knowledge gained from the tested, assessed, and matured prototype, a full-fledged and open-source version of the prototype with will be developed for testing by IFP developers and customers. The feedback of any deficiencies and unmet IFP requirements will be used to improve the benchmark tool. Also, the benchmark tool will be available as a licensed open-source software, so that others may use and extend it. Due to our DevOps Agile-driven methodology, we will establish a community of users associated with the benchmark tool at this point to ensure sustainability beyond the period of this project. The following research question should be answered in this phase: 

\textit{\textbf{RQ3a:} What would be the usability of the proposed benchmark tool from the viewpoints of both computing and non-computing users?} We will perform empirical assessments to measure how the benchmark tool is perceived by a diverse group of users and explore what improvements can be made from this experience. 

\textit{\textbf{RQ3b:} How would the benchmark tool help improve the existing IFPs?} We will perform an exploratory analysis from the use of the benchmark tool to find out the performance bottlenecks and vulnerabilities in existing representative IFPs including ICON, AION, and Wanchain. We will then develop and implement ideas to resolve the identified bottlenecks and eliminate the discovered security holes.  

The rest of the study is organized as follows. In Section 2, we discuss the state-of-the-art related work. Section 3 describes in detail the proposed research activities. Finally, Section 4 gives the concluding remarks.

\section{Related Work}
In this section, we discuss the studies, approaches and projects related to 1) blockchain interoperability and 2) blockchain benchmarking, as identified and proposed by both scientific and enterprise worlds.
\subsection{Blockchain Interoperability}
Blockchain interoperability is the idea of integration and interaction between multiple disparate blockchain systems (for instance, Bitcoin, Hyperledger Fabric, Ethereum etc.). Interoperable blockchains open doors to a world where moving assets and data from one blockchain to another are effortless and implementable without requiring any change in protocol of the base blockchain. In the following, we will review all related work in blockchain interoperability space and existing IFPs. 
In 2014, Back et al.\cite{23} were the first to propose the concept of \textit{2-way pegged sidechains} for the Bitcoin blockchain, which would allow transfer of Bitcoin and other assets from the Bitcoin blockchain to a sidechain and vice versa. Thus, users can have the flexibility to access various cryptocurrency systems by using the assets they already own. Sidechains are isolated from the main blockchain in such a way that in the case of a cryptographic break (or a maliciously designed sidechain), the damage is entirely confined to the sidechain itself.
 
Building up on the idea presented by Back et al.\cite{23}, Sztorc \cite{24} proposed \textit{Drivechains}, an SPV (Simplified Payment Verification) proof. As SPV proof relies on miner approval, this idea was based on using proofs to represent the approval directly: having miners “approve” these transactions in consecutive Bitcoin blocks (“miner vote” scheme). Although such a system might be space efficient and secure, these attributes are traded off by slow and infrequent asset withdrawals.

RSK \cite{25} is an evolution of QixCoin \cite{26}, a turing-complete cryptocurrency developed in 2013. RSK works as a Bitcoin Sidechain hence, when Bitcoins are transferred into the RSK blockchain, they become “\textit{SmartBitcoins}” (SBTC). \textit{SmartBitcoins} are equivalent to Bitcoins living in the RSK blockchain, and they can be transferred back to Bitcoins at any time for a standard RSK transaction fee. RSK provides federated pegs which can be a limitation in such systems as it reduces political decentralization, which is a key element in decentralized application along with architectural decentralization.

Interledger \cite{27} is a protocol for sending packets of money across different payment networks or ledgers. ILPv4 can be integrated with any type of ledger, including those not built for interoperability, and is designed to be used with a variety of higher-level protocols that implement features ranging from quoting to sending larger amounts of value with chunked payments. Although it sounds like a promising approach, the authors have not explicitly demonstrated the security aspects of the Interledger protocol.

Jin et al.\cite{28} proposed an architecture for supporting blockchain interoperability based on the five layers of the blockchain (application, contract, consensus, network, and data layer). Each layer implements a module to solve five interoperability challenges: guarantee of atomicity, efficiency improvement, security maintenance, diversification tolerance, and usability for developers.

Buterin \cite{29} presented a literature review and categorized the three primary strategies for blockchain interoperation: 1) \textit{Centralized or multi-signature notary schemes}, 2) \textit{Sidechains/relays}, and 3) \textit{Hash-locking}. The limitation of this study is that it does not take into consideration private or consortium chain settings. Such settings usually have their own unique challenges related to security and privacy, which the author deemed ‘out of scope’ of the study.

Culwick and Metcalf \cite{30} introduced Blocknet Design specification. Blocknet is a peer-to-peer protocol between nodes of different blockchains which enables the transfer of data and value between blockchains and opens the door to multi-chain architectures.

Using the Enterprise Ethereum Client Specification v1.0 \cite{31} as a basis, Robinson \cite{32} defined a set of additional requirements for establishing private Ethereum sidechains. The author also analyzed three blockchains namely Geth, Parity and Hyperledger Fabric  based on these combined requirements. The analysis showed that although these blockchain platforms comply with some basic requirements, they differed significantly and missed several requirements. The author did not discuss any requirements related to multi-chain communication and transactions.

Li et al. \cite{33} proposed a blockchain architecture which leverages the notion of \textit{satellite chains} that form interconnected, but independent, sub-chains, of a single blockchain system. Nodes can join a given satellite chain if they want to transact with another set of nodes. Each satellite chain maintains its own private ledger, thus preventing any non-member node from receiving or accessing any given transaction in its ledger.

Herlihy \cite{34} provided a protocol for Atomic cross-chain swaps. The protocol proved that: 1) if all parties conform to the protocol, then all swaps take place, 2) if some coalition deviates from the protocol, then no conforming party ends up worse off, and 3) no coalition has an incentive to deviate from the protocol.

There are various IFP platforms such as ICON \cite{11}, AION \cite{10} and Wanchain \cite{12} that have been proposed to solve the blockchain interoperability issue. Table \ref{tt1} presents a high-level comparison of these current IFPs based on their platform design and interoperability attributes.
\begin{table}[]
\centering
	\caption{Comparison between ICON, AION and Wanchain IFPs.(*LFT - Loop Fault Tolerance, PoW - Proof of Work, DPoS - Delegated Proof of Stake, PoI - Proof of Identity, PoS - Proof of Stake,
		DApps - Decentralized Applications.)}
	\label{tt1}
	\begin{tabular}{|l|l|l|l|l|}
		\hline
		General Attributes                & Specific Attributes & ICON & AION             & Wanchain \\ \hline
		\multirow{3}{*}{Platform Design}  & Virtual Machine     & No   & Yes              & Yes      \\ \cline{2-5} 
		& Consensus*          & LFT  & PoW + DPoS + PoI & PoS      \\ \cline{2-5} 
		& DApps*              & Yes  & Yes              & Yes      \\ \hline
		\multirow{4}{*}{Interoperability} & Bridging Protocol   & No   & Yes              & No       \\ \cline{2-5} 
		& Transfer of Value   & Yes  & Yes              & Yes      \\ \cline{2-5} 
		& Transfer of Logic   & No   & Yes              & No       \\ \cline{2-5} 
		& Interchain DApps    & No   & Yes              & No       \\ \hline
	\end{tabular}
\end{table}

\subsection{Blockchain Benchmarking}
Research in computer and network systems relies on open source benchmarking tools to evaluate systems and perform repeatable empirical studies. The blockchain domain has just surpassed its first decade after emergence so there are a very limited number of tools for assessing blockchain systems. Even worse, there are currently no IFP benchmark tools for the community to support research and development in the multi-blockchain paradigm. In this section, we discuss all blockchain-related tools and approaches discussed in literature and industrial practice, which provide some kind of evaluation and benchmarking support.

Dinh et al. \cite{14} proposed BLOCKBENCH, a benchmarking framework for evaluating private blockchains. This framework provides workloads for measuring the data processing performance, and workloads for understanding the performance at different layers of a single  blockchain. They implemented their benchmarks  by writing smart contracts to be integrated to the blockchain under study. BLOCKBENCH measures overall and component-wise performance in terms of throughput, latency, scalability and fault-tolerance.

Kalodner et al. \cite{35} presented \textit{BlockSci}, an open-source software for single-level blockchain analysis. It incorporated an in-memory analytical database and supports Bitcoin, Litecoin, Namecoin and Zeash blockchains. The limitation of BlockSci is that it does not support smart contracts blockchain platforms such as Ethereum.

Gervais et al. \cite{36} presented a framework to objectively compare PoW (Proof of Work) blockchains. Their framework can evaluate the impact of network-layer parameters on the security of PoW-based blockchain. With their analysis they found that Ethereum needs at least 37 block confirmations in order to match Bitcoin’s security with 6 block confirmations, given an adversary with 30\% of the total mining power. The results indirectly suggest that Bitcoin’s blockchain offers more security than Ethereum’s blockchain.  

Croman et al. \cite{37} presented experimental measurements of a range of metrics that characterize the resource costs and performance of today’s operational Bitcoin network. Their findings suggest that fundamental protocol redesign is needed for blockchains to scale significantly while retaining its decentralization. They have proposed the design of new protocols by partitioning blockchain systems into distinct planes, namely: Network, Consensus, Storage, View, and Side Planes.

Eyal et al. \cite{38} proposed a new protocol for the scaling of the Bitcoin blockchain. Additionally, they introduced several metrics for quantifying security and efficiency of Bitcoin-like blockchain protocols. They implemented Bitcoin-NG and performed large-scale experiments at 15\% the size of the operational Bitcoin system, using unchanged clients of both protocols. The experiments show that Bitcoin-NG scales optimally with bandwidth (but is limited by the capacity of the individual nodes) and latency (which is also limited only by the propagation time of the network).

In Ethereum blockchain, various aspects including block processing time, and transactions processing time, have also been benchmarked in experiments and studies. Kiffer et al. \cite{39} developed a simple Markov-chain based method for analyzing consistency properties of blockchain protocols. The method includes a formal way of stating strong concentration bounds as well as effortless ways to concretely compute the bounds.

Implementing a universal blockchain that could fit all use cases is not feasible to achieve with the current methodologies. Thus, it is practical to establish interoperability-facilitating platforms (IFPs) that could govern interoperability between disparate blockchains. Although research for blockchain interoperability has been growing, there are still no guidelines and tools available for the correct implementation and benchmarking of IFPs to render trade-off analysis and assessment for blockchain customers and developers. As the blockchain community takes strides towards a world with multi-blockchain ecosystem, new benchmarking tools that can analyze such interoperable systems is crucial.

\section{Proposed Research Activities}
This project's outcome will provide a comprehensive environment - a tool - for blockchain customers/users and IFP developers to perform benchmarking assessments on different IFPs in relation to multi-blockchain paradigm. The support comes as a result of three major research activities: IFP requirements specification and core benchmarking document/guidelines (see Section 3.1), extensible benchmarking framework and a prototype (see Section 3.2), and a full-fledged tool, open-source version of the prototype (see Section 3.3).  Fig 1 depicts the overview of proposed workflow or task-flow.

We will start by thoroughly fleshing out the current IFPs designs and techniques. We will undertake an extensive study of the existing ecosystem of multi-blockchain interoperability to determine the specific needs of IFPs and the types of requirements that need to be supported, and also develop core benchmarks and their characteristics. In parallel, we will build a prototype system as a basis for benchmarking tool; and share the prototype to get feedback. We will also gradually iterate the process of refining the underlying design, build new features of the prototype for any unmet IFP requirements, and evaluate the tool’s effectiveness for meeting the needs of blockchain customers and IFP providers.

\begin{figure}  
	\centering
	\includegraphics[scale=0.18]{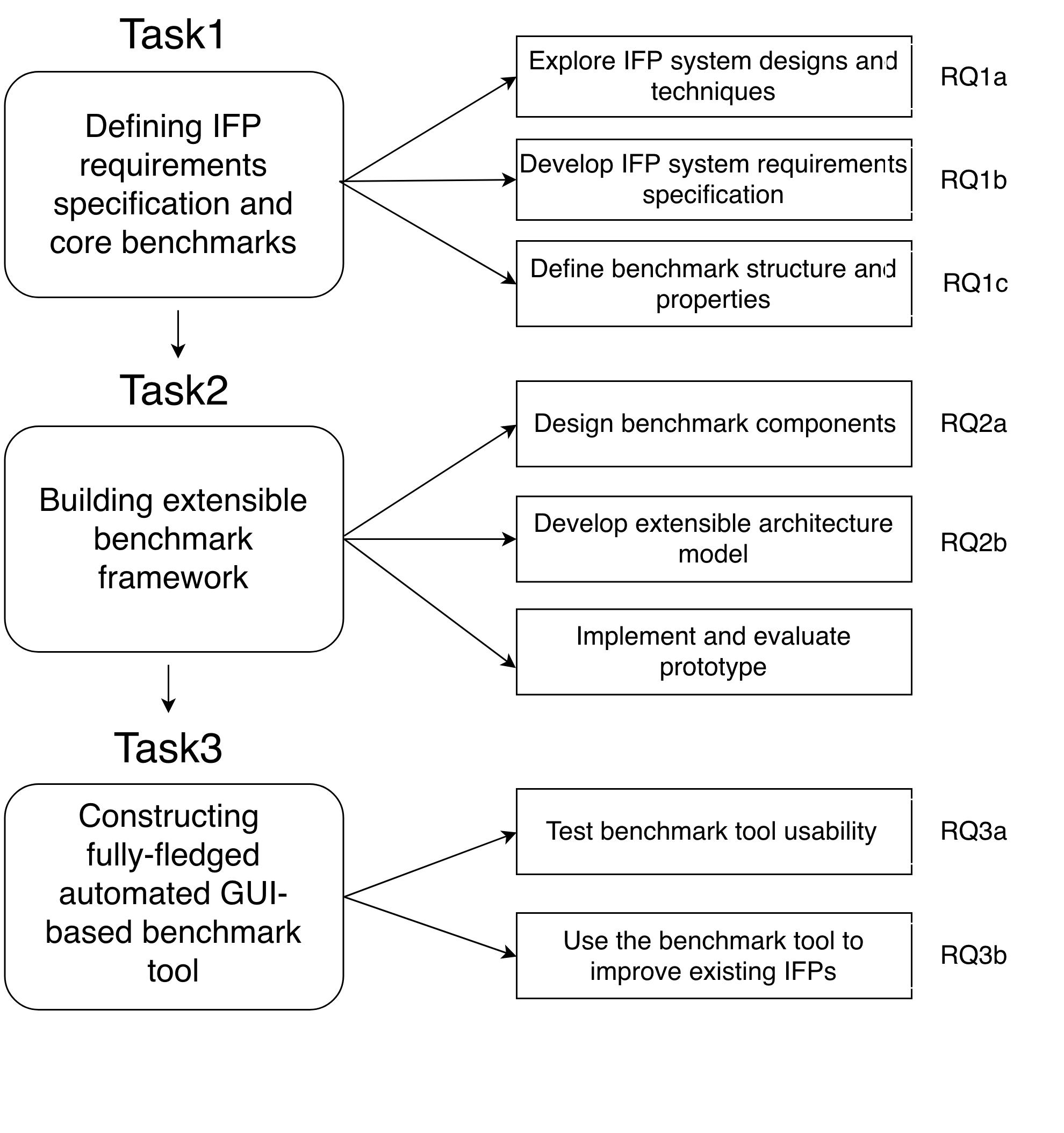}	
	{\scriptsize	\caption{ Overview of proposed work }}
\end{figure}

\subsection{Task 1: Defining IFP requirements specification and core benchmarks}
As mentioned earlier, the goal of this task is to understand the inner workings of existing IFPs by conducting an extensive study of the state-of-the-art work (see Section 3.1.1). We will then use the insights gained from the study to develop IFP requirements specification (see Section 3.1.2). Furthermore, we will define the properties and structuralize the core benchmarks that work towards the representative IFPs, such as ICON, AION, and Wanchain (see Section 3.1.3). 
\subsubsection{Explore IFP system designs and techniques in practice and theory (RQ1a)}
Nowadays, an increasing number of IFPs have been proposed or implemented by startups and researchers \cite{10, 11, 12}. All of them have the goal of connecting different blockchains to enable cross-chain transactions, but their designs and internal mechanisms convergent. For example, Wanchain is an Ethereum based IFP which supports transactions between permissionless and permission blockchains. This IFP focuses on crypto assets transfer and the cross-chain transactions in terms of value exchange. It uses cryptographic technologies such as secure multi-party computation \cite{a5} and ring signatures to ensure the privacy of the cross-chain transactions. While in AION, a transaction issued by the source blockchain will be wrapped with routing information so that the transaction can be delivered to the destination blockchain. At the same time, BTC Relay is another IFP that only focus on bridging the Bitcoin blockchain \cite{3} and the Ethereum smart contracts. 

In this part of the project, we will do a comprehensive study of the existing IFPs. On one hand, we will carefully go over the related documents, such as white paper and research papers, to understand the details of each IFP. This will include its targeted customers, system architecture, transaction workflow, consensus algorithm, reliability, and applicability. Analysis will be carried out to investigate the advantages and limitations of each IFP and compare different IFPs. Since most of these IFP projects are open-sourced, we will run different IFPs to understand their usability and performance. Such hands-on experiences will help us gain valuable insight into how different IFPs work, as well as strengths and weaknesses. 

\subsubsection{Develop IFP system requirements specification (RQ1b)}
Based on the observations we gain from the preceding study (as discussed in Section 3.1.1), we will develop requirements specification for IFP with focus on high-performance and secure interoperability constraints between different blockchains. Some Ethereum related specifications have been developed to set standards for developing large-scale blockchain implementations. For example, the Enterprise Ethereum Alliance (EEA) created the Enterprise Ethereum Client Specification V2 to give businesses and developers a comprehensive and instantly accessible view of the enterprise environment using Ethereum. Furthermore, Peter Robinson \cite{32} proposed the specifications for Ethereum Private Sidechains. Robinson also provided additional requirements for the Enterprise Ethereum Client Specification for API access control and data privacy. Inspired by our understanding of the current IFP systems and  the existing blockchain related specifications, we will develop IFP requirements specification from multiple categories. 

The first category is \textit{performance related specification}, which can be used for guidelines to design a high performance IFP. The specifications in this category primarily tackle the architecture of the IFP that can support high throughput and low latency cross-chain transactions \cite{a6}. The second category is security related, which will provide suggestions on how to make the cross-chain transactions reliable and private. Further, this category of specs includes IFP properties that are required to survive security attacks, including both traditional, e.g., Deny Of Service attack \cite{40, a7} and blockchain-specific attacks, e.g., Sybli attack \cite{41}. The third category is related to the usability of the IFP. These specifications will focus on guidelines to create an easy to use IFP system. Example scenarios of questions that will drive the development of this class of specs include: What APIs should an IFP provide to the client for cross-chain transaction invocation or query? What level of details should an IFP expose to the client in terms of cross-chain transaction control? and, How could an IFP be designed to be transparent enough to allow source and destination blockchains to be easily connected? 

The requirements specification developed in this phase will be further divided into two types - mandatary and suggested. Mandatory specifications are the more fundamental requirements that an IFP must meet in order to guarantee its correctness and reliability, while suggested ones could be customized based on user demands. 
\subsubsection{Define Benchmark structure and properties (RQ1c)}
In order to carry out a thorough measurement of the IFPs, the benchmark tool will have multiple tiers; with each tier focusing on a different set of metrics of the IFPs. The first tier is related to the IFP \textit{functionality} and aims to verify the correctness of the cross-chain transactions on the IFP.  The first tier also ensures there will be no inconsistency introduced between the source blockchain and the destination blockchain. The second tier is \textit{performance-related}. This tier focuses on the performance-related characteristics of the IFPs, addressing such issues as the time it takes for a cross-chain transaction to settle down; the maximum throughput that an IFP can achieve under different workloads; the throughput of an IFP; and scalability, with an increasing number of blockchains and cross-chain transactions connected to it. This second tier will also profile the resource utilization of an IFP, such as CPU usage, memory usage, and network usage. Such performance-related characteristics will, on one hand, help customers understand which IFP will provide better performance for their specific workloads. On the other hand, such characteristics will also help IFP providers optimize their infrastructure and minimize costs. The third tier is related to IFP\textit{ security}. Designed to support simulation of different attacks, the benchmark tool benefit the evaluation of an IFP’s perform reliability in the existence of compromised or malicious customers. Since an IFP is usually shared by multiple customers, this feature is critical to prevent attacks in one customer blockchain which might affect or infect other customer blockchains.

\subsection{Task 2: Building the benchmarking framework with extensible architecture}
A framework (system architecture) will be designed and built to serve as the backbone of the benchmark tool for generating the client workloads based on user specifications, defining new benchmarks, and executing benchmarks on IPFs. The primary design goal of the framework is extensibility. Another of the goals is to make it easy for both blockchain customers and developers to benchmark the increasing variety of IFP systems. The basic components of the framework are discussed in Section 3.2.1. The architectural model of the framework is discussed in Section 3.2.2. Section 3.2.3 is on the plan for the prototype development and initial assessment. 
	
\subsubsection{Design benchmark components (RQ2a)}
To achieve this subtask, we will design five major components as the building blocks of the benchmarking framework. The descriptions of these components are discussed as follows:

1) \textit{Client Workload Generator (CWG):} The client workload generator allows a client to generate workloads based on selected properties and parameters (discussed in Section 3.2.2) for the execution on a target IFP. So far, three core workloads have been identified, and are discussed below. We intend to develop and implement additional workloads for benchmarking and assessment throughout the IFP requirements specifications gathering process described in Task 1.

\begin{enumerate}[a.]
	\item
		\textit{No\_Action}: We will design a program that would accept a cross-chain transaction that simply “returns” to its callee. The program would allow a client to analyze performance at the consensus level as there are minimal operations required during the execution of these workloads.   Performance could directly be measured merely by analyzing the time to reach consensus or transaction latency.
	\item
		\textit{Cross-chain Transaction Processing (CTP)}: We will design a program to generate workloads for performing simple cross-chain transfer operations, for instance, the transfer of  assets from one account on the source blockchain to another account on the destination blockchain.
	\item
		\textit{ReadWrite\_Extreme (RWE)}: This module invokes a program to perform a substantial number of random cross-chain ‘read’ and ‘write’ operations through the target IFP. This would allow the client to estimate the IFP throughput, the transaction latency, and the disk and network I/O efficiency of the IFP.  
\end{enumerate}

2) \textit{Client Workload Executor (CWE)}: The client workload executor component contains the code to load and execute the workload generated by the \textit{Client Workload Generator} on a target IFP. The CWE component executes the client workloads by making calls to the IFP interface layer. The benchmark results will be collected during the workloads execution by the benchmarking tool, specifically by the \textit{Transaction Behavior Monitor} component (see below). Any run-time data from the CWE will be then fed to the statistics module, which consists of the \textit{Visualized Results Reporter} component (see below).

3) \textit{Transaction Behavior Monitor (TBM)}: The transaction behavior monitor records and captures the benchmark statistics for a target IFP during the workload execution in real time. Additionally, this component will serve as the data source for the client to simulate how the target IFP performs during the execution of the performance testing workloads.

4) \textit{IFP Interface Layer (IIL)}: This component serves as an abstract layer of the IFP and connects the benchmark tool to the IFP. Thus, this component translates requests from client threads into calls against the IFP. It is responsible for carrying out operations based on the generated workload, executing it by sending transactions for querying the state of the target IFP.  The benchmark tool also collects the IFP specific benchmark metrics such as consensus efficiency and resource utilization via the IFP interface layer.

5)\textit{ Visualized Results Reporter (VSR)}: To visualize the resulting data from the tool, the VSR will be used to convert the extracted benchmark results (from the TBM) of a target IFP into human readable visualization graphs and charts, including but not limited to, bar graphs, pie charts, and histograms.

\subsubsection{Develop architecture model (RQ2b)}
The framework will be designed with extensibility and scalability in mind. We will take great care to design a framework which can be integrated into new IFPs with relative ease. Many of the anticipated changes to the system will only require adding new types of IFP data and changing the user-defined benchmarks to make use of the system. Thus, the underlying architecture of the framework will only require “plugging in” these new types of data without refactoring the logic that passes the data over the network, retrieves and updates the results. The extensibility requirement will be implemented using well-known design patterns \cite{42}. Figure 2 depicts the high-level (generic) architectural model of the framework.

\begin{figure}  
	\centering
	\includegraphics[scale=0.55]{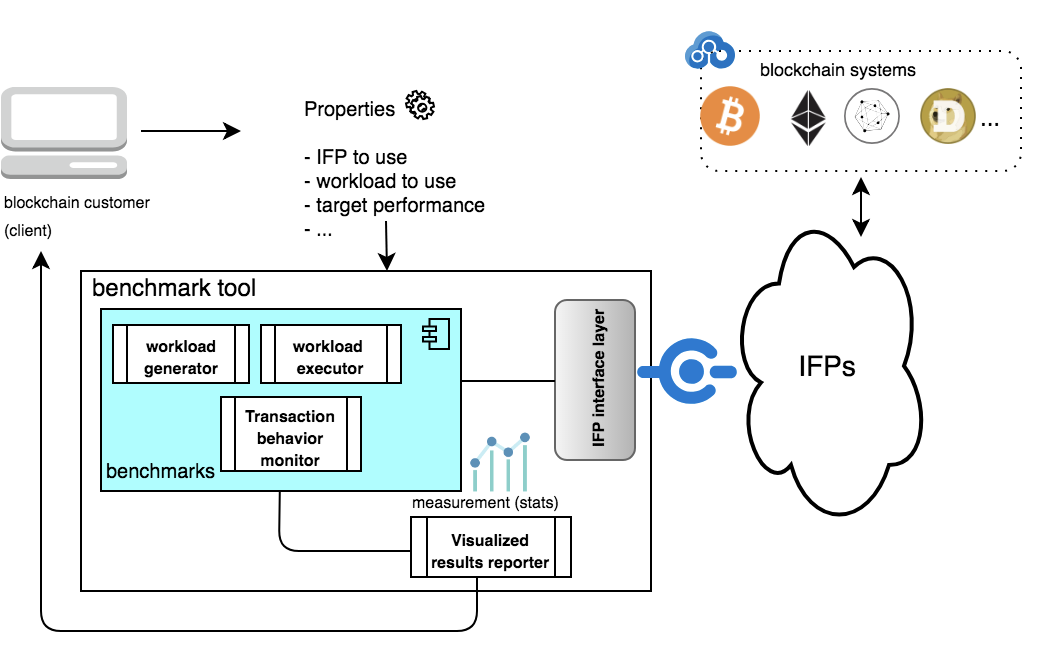}	
	{\scriptsize	\caption{ Conceptual architecture model of the framework }}
\end{figure}

As depicted in Figure 2, a typical client takes a series of properties which define the target IFP. The properties are divided into two categories: a) workload properties (those that define the client workloads independent of a given IFP or experimental run) and b) runtime properties (those specific for an IFP in use, and properties used to configure the IFP interface layer, among others). 

The basic flow, will be that the workload generator component builds the client workloads according to the user-specified workload properties, as shown in Figure 2. The workload executor component then drives client workloads by invoking the IFP interface layer to load data and execute the workload (the transaction phase in blockchains). The transaction behavior monitor component measures and collects the performance (e.g., latency and throughput), and security metrics (e.g., vulnerability-attack tolerance), then forwards these measurements to the statistics module. The visualized results-reporter component then converts benchmark data/results  into human friendly visuals including, but not limited to, bar graphs, pie charts, histograms, etc., and reports them back to the client.

We will use the microservices architecture \cite{22} as part of our overall design methodology to build the framework as a set of cohesive services. By using this architecture, components can be easily updated and replaced. This architecture not only allows the extensibility that we desire but also provide a user-friendly capability for blockchain customers as microservices are built around business capabilities where each service is scoped to a single purpose. In addition, this architecture gives to the users enriched implementation choices, from different programming languages, to write microservices and deploy them independently, as a single service, or as a group of services. In this way, each service can run in its own process and communicate with other services through a well-defined interface using a lightweight mechanism, typically an HTTP-based API. 

\subsubsection{Implement prototype and initial assessment }
A prototype system will be developed in this stage as a command-line program with semi-automated functionality. All prototype related artifacts (including well-defined set of IFP requirements specification, core benchmarks, design architecture, and framework details) will be hosted on GitHub, serving as the development site for researchers and developers to use and extend. In addition to the GitHub site, we will develop a project Website platform to support the tool for long-term community engagement and continuous improvement. The platform will provide a wide range of services including educational materials, tutorial videos and user guide, user questionnaire, news and releases, assessment results reports, to name a few.

To accompany the prototype, we will also publish our own assessment results on the commonly used IFPs, including ICON, AION, and Wanchain by running the prototype on the proposed core benchmarks. This assessment will help gain insights into the functionality of the benchmark tool, and consequently help its reliable evolution through the life of the project. In consideration of  user feedback and results, we will encourage both IFP developers and blockchain customers (i.e., the user communities) to: (1) publish and compare their results with our initial assessment results, and (2)  report any new results with emerging IFPs other than commonly used ones which have been included in the initial assessment. In order to ensure a fair, sustainable, and community-wide validation, we will ask users to use the following guidelines when publishing results (with emphasis on results that are easily reproducible):

\begin{itemize}
	\item {A description of the default out-of-the-box installation, including version and build numbers. }
	\item {A description of the IFP under evaluation and its connected blockchain systems.}
	\item{Decreasing complexity and process time of model with using the multi-class feature and parallelizing in ensemble algorithm.}
	
	\item {All configuration steps, tailoring, and onboarding that were performed to make the prototype tool run.}
	
	\item {All changes to default settings, benchmarks, or services used to achieve the results.} 
	
\end{itemize}

\subsection{Task 3: Constructing the full-fledged automated GUI-based IFP benchmark tool }
The major objective of this task is to extend the prototype with a user-friendly GUI and fully automated functionality. Corresponding to Task 3 research questions, the usability evaluation process of the benchmark tool is discussed in Section 3.3.1, which also includes details about our plan to extend the prototype in terms of new benchmark instantiation and the GUI. Section 3.3.2 explains our exploratory analysis plan for further improvements of IFPs, informed by the usage data from the initial prototype and the benchmark tool.
	 
\subsubsection{Conduct usability evaluation (RQ3a)}
Before conducting the usability evaluation, we will extend the prototype with a user-friendly GUI which would allow customers to define new benchmarks and carry out automated experiments. Afterwards, we will empirically assess the usability of our tool from viewpoints of both computing and non-computing users. In this way, we would able to measure how the benchmark tool is perceived by a diverse group of users and explore what improvements can be made from this experience. Our hypothesis is that having such an assessment done by external users of different backgrounds will increase the likelihood of any requirements that were not implemented in the benchmark tool and allow modifications and further improvements. In addition, such a diverse group of participants will allow the research team to have a different perspective of the benchmark tool, which could lead to the creation of a bridge between various research areas within the university. 

\subsubsection{Explore opportunities for improving IFPs (RQ3b)}
We expect the benchmark tool to profoundly benefit blockchain customers as it will allow them to select the most suited IFP based on their particular needs. In this particular sub-task, we will perform an exploratory analysis on the collected data from the use of the benchmark tool, either by our research team (faculty and students) or the community (developers and commercial users). This data analysis is expected to help us to identify any weaknesses in the defined properties (functionality, performance, or security related) of the IFPs. The expected results from such analysis will be used to resolve the identified bottlenecks or weaknesses.

\section{Concluding remarks }

Benchmarking is recognized as one of the most effective means to evaluate the performance of a system under certain conditions, relative to the performance of another system in the field of computing and computer networking; and also, in measuring effectiveness in the transfer of system-relevant knowledge between customers and developers. This project proposal outlines a comprehensive research design to provide benchmark support for the performance and stability of the multi-blockchain paradigm, and facilitate blockchain interoperability applications.

\bibliographystyle{unsrt}
\bibliography{ref5}

\end{document}